\documentclass[prd,twocolumn,showpacs,preprintnumbers,nofootinbib]{revtex4}
\usepackage[dvips]{graphicx}
\usepackage{enumerate}
\usepackage{amsmath,amssymb}
\usepackage{mathrsfs}
\usepackage[dvips]{graphicx}
\usepackage{bm}

\begin{document}
\preprint{CHIBA-EP-195, 2012}

\title{
A nonperturbative construction of massive Yang-Mills fields without the Higgs field
}

\author{Kei-Ichi Kondo$^{1}$}

\affiliation{$^1$Department of Physics,  
Graduate School of Science, 
Chiba University, Chiba 263-8522, Japan
}
\begin{abstract}
In order to understand the so-called decoupling solution for gluon and ghost propagators in QCD,  
we give a nonperturbative construction of a massive vector field describing a  non-Abelian massive  spin-one particle, which has the correct physical degrees of freedom and is invariant under a modified Becchi-Rouet-Stora-Tyutin transformation, in a massive Yang-Mills model without the Higgs field, i.e., the Curci-Ferrari model. 
The resulting non-Abelian massive vector boson field is written using a  nonlinear but local transformation from the original fields in the Curci-Ferrari model. 
As an application, we write down a local mass term for the Yang-Mills field and a dimension-two condensate, which are exactly invariant under the modified Becchi-Rouet-Stora-Tyutin transformation,  Lorentz transformation and  color rotation.

\end{abstract}


\maketitle

\section{Introduction}

In this paper we consider a massive Yang-Mills theory \cite{YM54} without the Higgs field \cite{Higgs66}.  
A motivation  of this research stems from  some nonperturbative phenomena caused by strong interactions.
\begin{itemize}
\item[(i)]
Confinement and Green functions--- The deep infrared behaviors of the gluon and ghost Green functions are believed to be intimately connected to color confinement in QCD \cite{KO79,Gribov78}.  
In the Landau gauge, the decoupling solution \cite{decoupling,FMP09,BGP10}  for the gluon and ghost propagators is currently supported rather than the scaling solution \cite{scaling} by recent numerical simulations on large lattices in three and four spacetime dimensions \cite{decoupling-lattice}.
Quite recently, it has been shown \cite{TW11} that the decoupling solution for the gluon and ghost propagators can be well reproduced from a low-energy effective model of a massive Yang-Mills theory, which is a special case of the Curci-Ferrari (CF) model \cite{CF76}.
This feature is not restricted to the Landau gauge and is common to  manifestly Lorentz covariant gauges, e.g., the maximal Abelian gauge \cite{MCM06}, as pointed out and demonstrated in \cite{Kondo11}.
We can ask how color confinement in QCD is understood from the CF model.

\item[(ii)]
Glueball mass spectrum--- 
A glueball should be constructed from the fundamental degrees of freedom of QCD, i.e., quark, gluon and ghost. 
For instance, the potential model of \cite{Cornwall82} identifies   glueballs with bound states of massive gluons.  They are described simply by introducing a naive mass term for gluons, 
$\frac12 M^2 \mathscr{A}_\mu \cdot \mathscr{A}^\mu$, which however breaks the Becchi-Rouet-Stora-Tyutin (BRST) symmetry. 
We ask how we can introduce a BRST-invariant mass term for gluons to establish a firm field theoretical foundation for treating glueballs, which will enable us to answer how precisely the mass and spin of the resulting glueballs are related to those of the constituent gluons. 

\item[(iii)]
Vacuum condensates---  
Besides  gauge-invariant vacuum condensates represented by   $\langle \bar \psi \psi \rangle$ with mass dimension-three and $\langle \mathscr{F}_{\mu\nu}^2 \rangle $ with mass dimension four, which are very important to characterize the nonperturbative vacuum of QCD, there might exist an extra dimension two condensate.  In fact, such a lower dimensional vacuum condensate is needed from the phenomenological point of view.  
However, such a condensate cannot be constructed from  gauge-invariant local composite operators in the framework of the local field theory. 
A  BRST-invariant vacuum condensate of mass dimension two has been  constructed  in  \cite{Kondo01,KMSI02}.  However, it is just  on-shell BRST invariant. 
Can we construct an  off-shell  BRST invariant version of vacuum condensate of mass dimension two?

\end{itemize}

Another motivation of studying the CF model comes from the field theoretical interest, since the massive Yang-Mills theory without the Higgs field has an unsatisfactory aspect   as a quantum field theory.
Renormalizability \cite{tHooft71,tHooft71b} is an important criterion for a quantum field theory to be a calculable and predictable theory. 
In addition, physical unitarity \cite{tHooft71,tHooft71b,KO78,DV70,SF70,Boulware70,CF76b,Ojima82,BSNW96,KG67,DTT88,RRA04} is another important criterion for a quantum theory of gauge fields to be a meaning theory, which prevents unphysical particles from being observed.

In view of this, we remind the readers of the well-known facts:

\begin{itemize}
\item[(i)]
 The massless Yang-Mills theory satisfies both   renormalizability and  physical unitarity \cite{tHooft71,KO78}.

\item[(ii)]
 The massive Yang-Mills theory in which local gauge invariance is spontaneously broken by the Higgs field and the gauge field acquires the mass through the Higgs mechanism satisfies both   renormalizability and   physical unitarity 
\cite{tHooft71b}.
\end{itemize}
In fact, the unified theory of Glashow-Weinberg-Salam for the electromagnetic and weak interactions based on the spontaneous symmetry breaking: $SU(2)_L \times U(1)_Y \rightarrow U(1)_{EM}$ predicted the massive gauge bosons $W^+, W^-$, and $Z^0$ which have been discovered in the mid-1980s, and the remaining Higgs particle is about to be discovered. 

However, in all the models proposed so far as the massive Yang-Mills theory without the Higgs fields (in which the local gauge symmetry is not spontaneously broken), it seems that    renormalizability and   physical unitarity are not compatible with each other.
See \cite{DTT88,RRA04} for reviews and \cite{BFQ} for later developments.
Indeed, the CF model has been shown to be renormalizable \cite{CF76b, BSNW96}, whereas  the CF model does not seem to satisfy   physical unitarity according to \cite{CF76b,Ojima82,BSNW96}.
Although the CF model is not invariant under the usual BRST transformation, it can be made invariant by modifying the  BRST transformation. But, the modified BRST transformation is not nilpotent.

It is known that   nilpotency is the key property to show   physical unitarity in the usual massless Yang-Mills theory,  
since the unphysical states form the BRST quartets and the cancellations occur among the quartets (Kugo-Ojima quartet mechanism) \cite{KO79,KO78}. 
It is not so clear if  nilpotency is necessary to recover physical unitarity in the massive case.
The physical unitarity of the CF model will be discussed in the perturbative and a  nonperturbative framework   in forthcoming papers \cite{Kondo12}.



Finally, it is instructive to mention the Gribov-Zwanziger (GZ) model \cite{GZ} and its modified version called the refined Gribov-Zwanziger (rGZ) model \cite{rGZ}, in comparison with the CF model, since GZ and rGZ  have been extensively studied in recent years to understand the scaling solution and the decoupling solution respectively. 
Common features to both models are as follows. 
\begin{itemize}
\item[(i)]
Both models have a dimensionful parameter to be determined afterwards, i.e., the ``mass'' parameter $M$ in the CF model and the Gribov parameter $\gamma$ in the rGZ model. 

\item[(ii)]
Both models are multiplicatively renormalizable to all orders of perturbation theory. 

\item[(iii)]
Both models do not respect the BRST symmetry: 
The Lagrangian is not invariant under the BRST transformation.

\item[(iv)]
Both models do not have the proof of fulfilling physical unitarity.

\end{itemize}

An advantage  of the CF model over  the rGZ model is that the CF model is much simpler than the rGZ model in the following sense:
 
\begin{itemize}
\item[(i)]
The CF model has the same field contents as those in the original Yang-Mills theory, while the rGZ model has additional unfamiliar fields introduced to rewrite the original nonlocal GZ model into a local field theory. 

\item[(ii)]
The CF model can explain the decoupling solution using one parameter $M$, while the rGZ model needs one more parameter which is related to the dimension-two condensate of the extra field (called the Zwanziger ghost), which is a nonlocal quantity in the original theory. 

\item[(iii)]
The CF model has the modified BRST symmetry \cite{CF76} which remains a local symmetry and reduces to the usual BRST symmetry in the limit $M \to 0$, while the GZ model has differently modified BRST symmetries which become inevitably nonlocal \cite{Sorella09,Kondo09b}. 
 
\end{itemize}

This paper is organized as follows.
In sec. II, we obtain the modified BRST and anti-BRST transformation for the CF model.
Although the result was already known in the CF paper \cite{CF76}, we give a constructive derivation of the modified BRST transformation to see how the resulting modification of the BRST transformation is unique under certain conditions.  

In sec. III,  using the fact that the CF model respects color symmetry, we rewrite the field equation of the Yang-Mills field in the Maxwell-like form.  This is an important technical tool needed in the next section to construct the massive vector field.

In sec. IV, we construct a non-Abelian massive vector field describing a massive  spin-one particle, which has the correct physical degrees of freedom and is invariant under a modified BRST transformation, from the Curci-Ferrari model  without the Higgs field. 
The resulting massive vector boson field is written by using a  nonlinear but local transformation from the original fields in the Curci-Ferrari model. This is the main result. 
As an application, we write down the mass term for the Yang-Mills field and a dimension-two condensate, which are invariant simultaneously under the modified BRST transformation,  Lorentz transformation, and  color rotation.

In Appendix A, we give a proof that    the path-integration measure is invariant under the modified BRST transformation.
In Appendix B, we show that even the modified BRST (and anti-BRST) invariant quantity depends on a parameter $\beta$ in the $M \not= 0$ case.  
This should be compared with the  $M=0$  case, in which $\beta$ is a gauge-fixing parameter and the BRST-invariant quantity does not depend on $\beta$, which means that the physics does not depend on $\beta$ in the $M=0$ case. 
This is not the case for $M \not= 0$  \cite{Lavrov12}.

\section{The Curci-Ferrari model and the modified BRST symmetry}

As a candidate of the massive Yang-Mills theory without the Higgs field,
we start from the Lagrangian density of the usual massless Yang-Mills theory in the most general Lorentz gauge \cite{Baulieu85} plus the ``mass term'' $\mathscr{L}_m$:
\begin{subequations}
	\begin{align}
		\mathscr{L}^{\rm{tot}}_{m\rm{YM}} =& \mathscr{L}_{\rm{YM}} + \mathscr{L}_{\rm{GF+FP}} + \mathscr{L}_{m} , \nonumber\\
		\mathscr{L}_{\rm{YM}}  =& - \frac{1}{4} \mathscr{F}_{\mu \nu} \cdot \mathscr{F}^{\mu \nu} ,  \\
		\mathscr{L}_{\rm{GF+FP}}  =& \frac{\beta}{2} \mathscr{N} \cdot \mathscr{N} + \mathscr{N} \cdot \partial^{\mu} \mathscr{A}_{\mu} 
		- \frac{\beta}{2} g \mathscr{N} \cdot (i \bar{\mathscr{C}} \times \mathscr{C}) \nonumber\\
		& + i \bar{\mathscr{C}} \cdot \partial^{\mu} \mathscr{D}_{\mu}[\mathscr{A}] \mathscr{C}
		+ \frac{\beta}{4} g^2 (i \bar{\mathscr{C}} \times \mathscr{C}) \cdot (i \bar{\mathscr{C}} \times \mathscr{C}) \nonumber\\
		 =& \mathscr{N} \cdot \partial^{\mu} \mathscr{A}_{\mu} + i \bar{\mathscr{C}} \cdot \partial^{\mu} \mathscr{D}_{\mu}[\mathscr{A}] \mathscr{C} \nonumber\\
		 		& + \frac{\beta}{4} ( \bar{\mathscr{N}} \cdot \bar{\mathscr{N}} + \mathscr{N} \cdot \mathscr{N}),  \\
		\mathscr{L}_{m}  =& \frac{1}{2} M^2 \mathscr{A}_{\mu} \cdot \mathscr{A}^{\mu} + \beta M^2 i \bar{\mathscr{C}} \cdot \mathscr{C} , 
	\end{align}
	\end{subequations}
where $\beta$ is a parameter corresponding to a gauge-fixing (GF) parameter in the $M \rightarrow 0$ limit,%
\footnote{
We can add a term
$
\frac{\alpha}{2}  \mathscr{N} \cdot \mathscr{N}
$ with another gauge fixing parameter $\alpha$.
} 
$\mathscr{D}_{\mu}$ is the covariant derivative defined by
	\begin{align}
 \mathscr{D}_{\mu}[\mathscr{A}] \omega(x) 
:= \partial_{\mu} \omega(x) + g \mathscr{A}(x) \times \omega(x) ,
	\end{align}
 and $\bar{\mathscr{N}}$ is defined by
\begin{equation}
\bar{\mathscr{N}} :=-\mathscr{N}+gi\bar{\mathscr{C}} \times \mathscr{C} .
\end{equation}
Here  the total Lagrangian density is written in terms of the Yang-Mills field $\mathscr{A}_\mu$, the Faddeev-Popov   ghost field $\mathscr{C}$, antighost field $\bar{\mathscr{C}}$ and the Nakanishi-Lautrup (NL) field $\mathscr{N}$, if we use the terminology in the usual massless Yang-Mills theory.
This is the Curci-Ferrari (CF) model \cite{CF76}.
In the Abelian limit with  vanishing structure constants $f^{ABC}=0$, the Faddeev-Popov ghosts decouple and the CF model reduces to the Nakanishi model \cite{Nakanishi72}.

The original Yang-Mills Lagrangian $\mathscr{L}_{\rm{YM}}$ is invariant under the gauge transformation: 
\begin{equation}
 \delta   \mathscr{A}_{\mu}(x) = \mathscr{D}_{\mu}[\mathscr{A}] \omega(x) .
\end{equation}
We remember that   $\mathscr{L}_{\rm YM} + \mathscr{L}_{\rm GF+FP}$ 
is constructed so as to be invariant 
under both the usual BRST transformation: 
	\begin{align}
		\begin{cases}
			{\boldsymbol \delta}  \mathscr{A}_{\mu}(x) = \mathscr{D}_{\mu}[\mathscr{A}] \mathscr{C}(x)  \\
			{\boldsymbol \delta}  \mathscr{C}(x) = -\frac{g}{2} \mathscr{C}(x) \times \mathscr{C}(x)   \\
			{\boldsymbol \delta}  \bar{\mathscr{C}}(x) = i \mathscr{N}(x)   \\
			{\boldsymbol \delta}  \mathscr{N}(x) = 0  \\
		\end{cases} ,
		\label{BRST}
	\end{align}
and anti-BRST transformation: 
\begin{align}
\begin{cases}
			\bar{\boldsymbol \delta}  \mathscr{A}_{\mu}(x) = \mathscr{D}_{\mu}[\mathscr{A}] \bar{\mathscr{C}}(x) \\
			\bar{\boldsymbol \delta}  \bar{\mathscr{C}}(x) 
= -\frac{g}{2} \bar{\mathscr{C}}(x) \times \bar{\mathscr{C}}(x)  \\
			\bar{\boldsymbol \delta}   \mathscr{C}(x) = i \bar{\mathscr{N}}(x) \\
			\bar{\boldsymbol \delta}  \bar{\mathscr{N}}(x) = 0
\end{cases} 
 .
\end{align}
Indeed, it is checked that  both $\mathscr{L}_{\rm{YM}}$ and $\mathscr{L}_{\rm{GF+FP}}$ are invariant under the BRST and anti-BRST transformations:
\begin{align}
		{\boldsymbol \delta}  \mathscr{L}_{\rm{YM}} = 0 ,
\quad
{\boldsymbol \delta}  \mathscr{L}_{\rm{GF+FP}} = 0 ,
\\
		\bar{\boldsymbol \delta}  \mathscr{L}_{\rm{YM}} = 0 ,
\quad
\bar{\boldsymbol \delta}  \mathscr{L}_{\rm{GF+FP}} = 0 .
	\end{align}	
	
We try to introduce a mass term $\mathscr{L}_{m}$ so that the total Lagrangian $\mathscr{L}^{\rm tot}_{m\rm{YM}}$ remains invariant under the BRST transformation.
However, we will find that the total Lagrangian is no longer invariant under the usual BRST transformation,
once the ``mass term'' $\mathscr{L}_{m}$ is introduced into the Yang-Mills theory. Indeed, we observe for a specific choice $\beta=0$: 
	\begin{align}
		{\boldsymbol \delta} \mathscr{L}_{m}
		&= M^2 \partial^{\mu} \mathscr{C} \cdot \mathscr{A}_{\mu} \ne 0 ,
	\end{align}
where we have used $(\mathscr{A}_{\mu} \times \mathscr{A}^{\mu})^A=g^{\mu \nu}(\mathscr{A}_{\mu} \times \mathscr{A}_{\nu})^A
=g^{\mu \nu}f^{ABC}\mathscr{A}_{\mu}^B \mathscr{A}_{\nu}^C = 0$.

In what follows, therefore, we consider if the total Lagrangian can be made invariant by modifying the BRST transformation.
For this purpose, we reexamine the BRST invariance of the GF+FP term, and  we try to find such a modified BRST transformation ${\boldsymbol \delta}'$.

Suppose that the modified BRST transformation ${\boldsymbol \delta}'$ of $\mathscr{A}_{\mu}$ has the same form as the usual BRST transformation:
	\begin{equation}
		{\boldsymbol \delta}' \mathscr{A}_{\mu}(x) = \mathscr{D}_{\mu}[\mathscr{A}] \mathscr{C}(x) ,
		\label{mBRST0}
	\end{equation}
to guarantee  the invariance of the original Yang-Mills action:
 	\begin{equation}
{\boldsymbol \delta}'\mathscr{L}_{\rm{YM}}=0 .
	\end{equation}
In order to realize ${\boldsymbol \delta}' \mathscr{L}^{\rm tot}_{\rm mYM}=0$, 
we require 
	\begin{align}
		0 = {\boldsymbol \delta}' (\mathscr{L}_{\rm GF+FP} + \mathscr{L}_{m}) .
		\label{req}
	\end{align}	
	Here we cannot require the invariance of the respective part, $\mathscr{L}_{\rm GF+FP}$ or $\mathscr{L}_{m}$, since we observe below that the choice (\ref{mBRST0}) inevitably leads to  ${\boldsymbol \delta}' \mathscr{L}_{m} \ne 0$, which means   ${\boldsymbol \delta}' \mathscr{L}_{\rm GF+FP} \ne 0$ to guarantee ${\boldsymbol \delta}' \mathscr{L}^{\rm tot}_{\rm mYM}=0$. 
	
First, we consider the $\beta=0$ case for simplicity. Then we have
	\begin{align}
		{\boldsymbol \delta}' \mathscr{L}_{\rm GF+FP}  =& {\boldsymbol \delta}' (\mathscr{N} \cdot \partial^{\mu} \mathscr{A}_{\mu}
		+ i \bar{\mathscr{C}} \cdot \partial^{\mu} \mathscr{D}_{\mu}[\mathscr{A}] \mathscr{C}) \nonumber\\ 
		 =& {\boldsymbol \delta}' \mathscr{N} \cdot \partial^{\mu} \mathscr{A}_{\mu}
		+ \mathscr{N} \cdot \partial^{\mu} \mathscr{D}_{\mu}[\mathscr{A}] \mathscr{C}
		 \nonumber\\ 
		&+ i {\boldsymbol \delta}'\bar{\mathscr{C}} \cdot \partial^{\mu} \mathscr{D}_{\mu}[\mathscr{A}] \mathscr{C}- i \bar{\mathscr{C}} \cdot \partial^{\mu} {\boldsymbol \delta}' {\boldsymbol \delta}' \mathscr{A}_{\mu} ,
	\end{align}
while
	\begin{equation}
		{\boldsymbol \delta}' \mathscr{L}_{m}
		= M^2 \partial^{\mu} \mathscr{C} \cdot \mathscr{A}_{\mu}
		= - M^2 \mathscr{C} \cdot \partial^{\mu} \mathscr{A}_{\mu} 
\not= 0.
	\end{equation}

In order to determine the modified BRST transformation for other fields, we perform:
	\begin{align}
	&{\boldsymbol \delta}' (\mathscr{L}_{\rm GF+FP} + \mathscr{L}_{m})
\nonumber\\ 	 =& ({\boldsymbol \delta}' \mathscr{N} -M^2 \mathscr{C}) \cdot \partial^{\mu} \mathscr{A}_{\mu}
		+ (\mathscr{N} + i {\boldsymbol \delta}' \bar{\mathscr{C}}) \cdot \partial^{\mu} \mathscr{D}_{\mu}[\mathscr{A}] \mathscr{C} \nonumber\\ 
		& - i \bar{\mathscr{C}} \cdot \partial^{\mu} \mathscr{D}_{\mu}[\mathscr{A}] ({\boldsymbol \delta}'\mathscr{C} + \frac{g}{2} \mathscr{C} \times \mathscr{C}) ,
	\end{align}
where we have used:
	\begin{align}
		  {\boldsymbol \delta}' {\boldsymbol \delta}' \mathscr{A}_{\mu} &= {\boldsymbol \delta}' (\mathscr{D}_{\mu} \mathscr{C}) \nonumber\\
		&= {\boldsymbol \delta}' (\partial_{\mu} \mathscr{C} + g \mathscr{A}_{\mu} \times \mathscr{C}) \nonumber\\
		&= \partial_{\mu} {\boldsymbol \delta}' \mathscr{C} + g (\mathscr{D}_{\mu}   \mathscr{C}) \times \mathscr{C} 
		+ g \mathscr{A}_{\mu} \times {\boldsymbol \delta}' \mathscr{C} \nonumber\\
		&= \mathscr{D}_{\mu} {\boldsymbol \delta}' \mathscr{C} + \frac{g}{2}   (\mathscr{D}_{\mu} \mathscr{C} \times \mathscr{C}   
		+ \mathscr{C} \times \mathscr{D}_{\mu} \mathscr{C}) \nonumber\\
		&= \mathscr{D}_{\mu} \left[{\boldsymbol \delta}' \mathscr{C} + \frac{g}{2} (\mathscr{C} \times \mathscr{C}) \right] .
	\end{align}
The requirement (\ref{req}) is satisfied, if we adopt
	\begin{equation}
		{\boldsymbol \delta}' \mathscr{N} = M^2 \mathscr{C} , \quad
		{\boldsymbol \delta}' \bar{\mathscr{C}} = i \mathscr{N} , \quad
		{\boldsymbol \delta}'\mathscr{C} = - \frac{g}{2} (\mathscr{C} \times \mathscr{C}) .
	\end{equation}
Thus we have found a modified BRST transformation:
	\begin{align}
		\begin{cases}
			{\boldsymbol \delta}' \mathscr{A}_{\mu}(x) = \mathscr{D}_{\mu}[\mathscr{A}] \mathscr{C}(x) , \\
			{\boldsymbol \delta}' \mathscr{C}(x) = -\frac{g}{2} \mathscr{C}(x) \times \mathscr{C}(x) , \\
			{\boldsymbol \delta}' \bar{\mathscr{C}}(x) = i \mathscr{N}(x) , \\
			{\boldsymbol \delta}' \mathscr{N}(x) = M^2 \mathscr{C} , \\
		\end{cases}
		\label{mod-BRST}
	\end{align}
which deforms the BRST transformation of the NL field and reduces to the usual BRST transformation in the limit $M \to 0$.
It should be remarked that 
 	\begin{equation}
{\boldsymbol \delta}' \mathscr{L}_{m} \ne 0, 
\quad
{\boldsymbol \delta}' \mathscr{L}_{\rm GF+FP} \ne 0.
	\end{equation}

Similarly, the total action is shown to be invariant under a modified anti-BRST transformation $\bar{{\boldsymbol \delta}}'$
defined by 
	\begin{align}
		\begin{cases}
			\bar{{\boldsymbol \delta}}' \mathscr{A}_{\mu}(x) = \mathscr{D}_{\mu}[\mathscr{A}] \bar{\mathscr{C}}(x) , \\
			\bar{{\boldsymbol \delta}}' \bar{\mathscr{C}}(x) = -\frac{g}{2} \bar{\mathscr{C}}(x) \times \bar{\mathscr{C}}(x) , \\
			\bar{{\boldsymbol \delta}}' \mathscr{C}(x) = i \bar{\mathscr{N}}(x) , \\
			\bar{{\boldsymbol \delta}}' \bar{\mathscr{N}}(x) = - M^2 \bar{\mathscr{C}}(x) , 
		\end{cases}
		\label{mod-BRST2}
	\end{align}
which reduces to the usual anti-BRST transformation in the limit $M \to 0$.
It is sometimes useful to give another form:
\begin{align}
 {\boldsymbol \delta}' \mathscr{\bar N}(x) =& g \mathscr{\bar N}(x) \times \mathscr{C}(x) - M^2 \mathscr{C}(x) ,
\nonumber\\
\bar{{\boldsymbol \delta}}' \mathscr{N}(x) =& g \mathscr{N}(x) \times \bar{\mathscr{C}}(x) + M^2 \bar{\mathscr{C}}(x) .
\end{align}

However, the modified BRST transformation violates the nilpotency:
	\begin{align}
\begin{cases}
		{\boldsymbol \delta}' {\boldsymbol \delta}' \mathscr{A}_{\mu}(x) = 0 , \\ 
		{\boldsymbol \delta}' {\boldsymbol \delta}' \mathscr{C}(x) = 0 , \\ 
		{\boldsymbol \delta}' {\boldsymbol \delta}' \bar{\mathscr{C}}(x) = i {\boldsymbol \delta}' \mathscr{N}(x)
		= i M^2 \mathscr{C}(x) \ne 0 , \\ 
		{\boldsymbol \delta}' {\boldsymbol \delta}' \mathscr{N}(x) = M^2 {\boldsymbol \delta}' \mathscr{C}(x)
	=	- M^2 \frac{g}{2} \mathscr{C}(x) \times \mathscr{C}(x) \ne 0 .
\end{cases} 
	\end{align}
The nilpotency is violated also for the modified anti-BRST transformation:
	\begin{align}
\begin{cases}
		\bar{{\boldsymbol \delta}}' \bar{{\boldsymbol \delta}}' \mathscr{A}_{\mu}(x) =  0 , \\
		\bar{{\boldsymbol \delta}}' \bar{{\boldsymbol \delta}}' \bar{\mathscr{C}}(x) =  0 , \\
		\bar{{\boldsymbol \delta}}' \bar{{\boldsymbol \delta}}' \mathscr{C}(x) 
=  i \bar{{\boldsymbol \delta}}' \bar{\mathscr{N}}(x)
		=  - i M^2 \bar{\mathscr{C}}(x) \ne 0 , \\
		\bar{{\boldsymbol \delta}}' \bar{{\boldsymbol \delta}}' \bar{\mathscr{N}}(x) =  - M^2 \bar{{\boldsymbol \delta}}' \bar{\mathscr{C}}(x)
		=  M^2 \frac{g}{2} \bar{\mathscr{C}}(x) \times \bar{\mathscr{C}}(x) \ne 0.
\end{cases} 
	\end{align}
In the limit $M \to 0$, the modified BRST and anti-BRST transformations reduce  to the usual BRST and anti-BRST transformations and become nilpotent.

Moreover, it is checked that the modified BRST and modified anti-BRST transformations  no longer  anticommute in the $M \not=0$ case:
	\begin{align}
\begin{cases}
		\{ {\boldsymbol \delta}' , \bar{\boldsymbol \delta}' \} \mathscr{A}_{\mu}(x) = 0 , \\ 
		\{ {\boldsymbol \delta}' , \bar{\boldsymbol \delta}' \} \mathscr{C}(x) = -iM^2  \mathscr{C}(x) ,  \\ 
		\{ {\boldsymbol \delta}' , \bar{\boldsymbol \delta}' \} \bar{\mathscr{C}}(x) =  iM^2  \mathscr{\bar C}(x), \\ 
		\{ {\boldsymbol \delta}' , \bar{\boldsymbol \delta}' \} \mathscr{N}(x) =   0 .
\end{cases} 
	\end{align}	
In the limit $M \to 0$, the anticommutativity is recovered: $\{ {\boldsymbol \delta}' , \bar{\boldsymbol \delta}' \}\to 0$.

Next, we consider the $\beta \ne 0$ case.
By using the modified BRST transformation, $\mathscr{L}_{\rm GF+FP}$ is rewritten as
	\begin{align}
	&	\mathscr{L}_{\rm GF+FP} \nonumber\\
 =& - {\boldsymbol \delta} \left[i \bar{\mathscr{C}} \cdot \left(\partial^{\mu} \mathscr{A}_{\mu} + \frac{\beta}{2} \mathscr{N}
		- \frac{\beta}{4} gi \bar{\mathscr{C}} \times \mathscr{C}  \right) \right] \nonumber\\
		=& - {\boldsymbol \delta}' \left[i \bar{\mathscr{C}} \cdot \left(\partial^{\mu} \mathscr{A}_{\mu} + \frac{\beta}{2} \mathscr{N}
		- \frac{\beta}{4} gi \bar{\mathscr{C}} \times \mathscr{C}  \right) \right]
		\nonumber\\
&- i \bar{\mathscr{C}} \cdot \frac{\beta}{2} {\boldsymbol \delta}' \mathscr{N} \nonumber\\
		=& - {\boldsymbol \delta}' \left[i \bar{\mathscr{C}} \cdot \left(\partial^{\mu} \mathscr{A}_{\mu} + \frac{\beta}{2} \mathscr{N}
		- \frac{\beta}{4} gi \bar{\mathscr{C}} \times \mathscr{C}  \right) \right]
		\nonumber\\
&- \frac{\beta}{2} M^2 i \bar{\mathscr{C}} \cdot \mathscr{C} \nonumber\\
		=& i {\boldsymbol \delta}' \bar{{\boldsymbol \delta}}' \left(\frac{1}{2} \mathscr{A}^{\mu} \cdot \mathscr{A}_{\mu}
		+ \frac{\beta}{2} i \bar{\mathscr{C}} \cdot \mathscr{C} \right) - \frac{\beta}{2} M^2 i \bar{\mathscr{C}} \cdot \mathscr{C} .
	\end{align}
Using the fact that only the transformation of the NL field ${\boldsymbol \delta}' \mathscr{N}$
is modified  in the modified BRST transformation, we find 
	\begin{align}
		& {\boldsymbol \delta}' \mathscr{L}_{\rm GF+FP} \nonumber\\
&= 
 {\boldsymbol \delta}' \mathscr{N} \cdot \partial^{\mu} \mathscr{A}_{\mu}
+ \beta {\boldsymbol \delta}' \mathscr{N} \cdot \mathscr{N}
		- \frac{g}{2} \beta {\boldsymbol \delta}' \mathscr{N} \cdot (i \bar{\mathscr{C}} \times \mathscr{C}) \nonumber\\
		&=  M^2 \mathscr{C} \cdot \partial^{\mu} \mathscr{A}_{\mu}+ \beta M^2 \mathscr{C} \cdot \mathscr{N}   
		- \frac{g}{2} \beta M^2 \mathscr{C} \cdot (i \bar{\mathscr{C}} \times \mathscr{C}) \nonumber\\
		&= M^2 \mathscr{C} \cdot \partial^{\mu} \mathscr{A}_{\mu} + \beta M^2 \left[ \mathscr{N} \cdot \mathscr{C}
		- i \bar{\mathscr{C}} \cdot \frac{g}{2} (\mathscr{C} \times \mathscr{C}) \right] \nonumber\\
		&= - M^2 {\boldsymbol \delta}' \left(\frac{1}{2} \mathscr{A}_{\mu} \cdot \mathscr{A}^{\mu} \right)
		- \beta M^2 {\boldsymbol \delta}' (i \bar{\mathscr{C}} \cdot \mathscr{C}) 
 .
	\end{align}
Therefore, a simple choice of the $\beta$-dependent  mass term $\mathscr{L}_{m}$ satisfying ${\boldsymbol \delta}' (\mathscr{L}_{\rm GF+FP} + \mathscr{L}_{m})=0$
is indeed given by
	\begin{equation}
		\mathscr{L}_{m} = \frac{1}{2} M^2 \mathscr{A}_{\mu} \cdot \mathscr{A}^{\mu} + \beta M^2 i \bar{\mathscr{C}} \cdot \mathscr{C} .
	\end{equation}
	Moreover, the path-integral integration measure  
$\mathcal{D} \mathscr{A} \mathcal{D} \mathscr{C} \mathcal{D} \bar{\mathscr{C}} \mathcal{D} \mathscr{N}$
is invariant under the modified BRST transformation. 
Indeed, it is shown in Appendix A that the Jacobian associated to the change of integration variables $\Phi(x) \to \Phi'(x) =\Phi(x)+ \lambda {\boldsymbol \delta}' \Phi(x)$ for the integration measure is equal to one.

\section{Field equations and color symmetry}

The  field equations are obtained as follows:
	\begin{align}
		\frac{\delta \mathscr{L}^{\rm tot}}{\delta \mathscr{A}^{\mu}}
		=& \mathscr{D}^{\nu}[\mathscr{A}] \mathscr{F}_{\nu \mu} - \partial_{\mu} \mathscr{N} + gi \partial_{\mu} \bar{\mathscr{C}} \times \mathscr{C} \nonumber\\&
		+ M^2 \mathscr{A}_{\mu} + g J_{\mu} = 0 , \nonumber\\
		\frac{\delta \mathscr{L}^{\rm tot}}{\delta \mathscr{N}}
		=& \partial^{\mu} \mathscr{A}_{\mu} + \beta \mathscr{N} - \frac{\beta}{2} gi \bar{\mathscr{C}} \times \mathscr{C} = 0 , \nonumber\\
		\frac{\delta \mathscr{L}^{\rm tot}}{i\delta \bar{\mathscr{C}}}
		=& \partial^{\mu} \mathscr{D}_{\mu}[\mathscr{A}] \mathscr{C} - \frac{\beta}{2} g \bar{\mathscr{N}} \times \mathscr{C}
		+ \beta M^2 \mathscr{C} = 0 , \nonumber\\
		\frac{\delta \mathscr{L}^{\rm tot}}{\delta \mathscr{C}}
		=& - \mathscr{D}_{\mu}[\mathscr{A}] i \partial^{\mu} \bar{\mathscr{C}} - \frac{\beta}{2} g \bar{\mathscr{N}} \times i \bar{\mathscr{C}}
		- \beta M^2 i \bar{\mathscr{C}} = 0 ,
		\label{field-eq}
	\end{align}
where we have used the left derivative and defined the matter current $J_{\mu}$ by
	\begin{align}
		J_{\mu}^A := g^{-1} \frac{\partial \mathscr{L}_{\rm matter}}{\partial \mathscr{A}^{\mu A}}
= -i (T^A \varphi)_a \frac{\partial \mathscr{L}^{\rm tot}}{\partial (\partial^{\mu} \varphi_a)} .
	\end{align}

We observe that the total Lagrangian of the CF model is invariant under the (infinitesimal) \textbf{global gauge transformation} or \textbf{color rotation} defined by
	\begin{align}
		&\delta \Phi(x) := [\varepsilon^C i Q^C, \Phi(x)] = \varepsilon \times \Phi(x) , \nonumber\\
 & \quad {\rm for} \quad
		 \Phi=\mathscr{A}_{\mu}, \mathscr{N}, \mathscr{C} , \bar{\mathscr{C}} ,  \\
		&\delta \varphi(x) := [\varepsilon^C i Q^C, \varphi(x)] = - i \varepsilon \varphi(x) ,
	\end{align}
or 
	\begin{align}
		&\delta \Phi^A(x) = f^{ABC} \varepsilon^B \Phi^C(x) , \nonumber\\
		&\delta \varphi_a(x) = - i \varepsilon^A (T^A)_a^{\ b} \varphi_b(x) = - i \varepsilon^A (T^A \varphi)_a .
	\end{align}

The associated conserved Noether current $\mathscr{J}^{\mu}$ is obtained from:
	\begin{align}
	 	& \varepsilon \cdot \mathscr{J}^{\mu}_{\rm color} \nonumber\\ 
		 =& \delta \mathscr{A}_{\nu} \cdot \frac{\partial \mathscr{L}}{\partial (\partial_{\mu} \mathscr{A}_{\nu})}
		+ \delta \mathscr{C} \cdot \frac{\partial \mathscr{L}}{\partial (\partial_{\mu} \mathscr{C})}
		+ \delta \bar{\mathscr{C}} \cdot \frac{\partial \mathscr{L}}{\partial (\partial_{\mu} \bar{\mathscr{C}})}
		 \nonumber\\ &
		+ \delta \mathscr{N} \cdot \frac{\partial \mathscr{L}}{\partial (\partial_{\mu} \mathscr{N})}
		+ \delta \varphi \cdot \frac{\partial \mathscr{L}}{\partial (\partial_{\mu} \varphi)} \nonumber\\
		 =& (\varepsilon \times \mathscr{A}_{\nu}) \cdot \mathscr{F}^{\nu \mu} + (\varepsilon \times \mathscr{C}) \cdot (i \partial^{\mu}\bar{\mathscr{C}}) \nonumber\\
		&  
		+ (\varepsilon \times \bar{\mathscr{C}}) \cdot (-i \mathscr{D}^{\mu}[\mathscr{A}] \mathscr{C})  
+ (\varepsilon \times \mathscr{N}) \cdot (- \mathscr{A}^{\mu})  + J^{\mu} \cdot \varepsilon \nonumber\\
		 =& \varepsilon  \cdot (\mathscr{A}_{\nu} \times \mathscr{F}^{\nu \mu}) +  \varepsilon \cdot (\mathscr{C} \times i \partial^{\mu}\bar{\mathscr{C}}) 
		-  \varepsilon \cdot (i \bar{\mathscr{C}} \times \mathscr{D}^{\mu}[\mathscr{A}] \mathscr{C})  \nonumber\\
		&  +  \varepsilon \cdot (\mathscr{A}^{\mu} \times \mathscr{N})  + \varepsilon \cdot J^{\mu}   .
	\end{align}
Thus the Noether current $\mathscr{J}^{\mu}$ associated with the color symmetry which is conserved $\partial_{\mu} \mathscr{J}^{\mu}_{\rm color}=0$ is given by 
	\begin{align}
	&	\mathscr{J}^{\mu}_{\rm color} \nonumber\\ 
		 =& \mathscr{A}_{\nu} \times \mathscr{F}^{\nu \mu} + \mathscr{C} \times i \partial^{\mu}\bar{\mathscr{C}}
		- i \bar{\mathscr{C}} \times \mathscr{D}^{\mu}[\mathscr{A}] \mathscr{C} + \mathscr{A}^{\mu} \times \mathscr{N} \nonumber\\
		&+ J^{\mu} \nonumber\\
		 =& \mathscr{A}_{\nu} \times \mathscr{F}^{\nu \mu} + i \partial^{\mu}\bar{\mathscr{C}} \times \mathscr{C}
		- \mathscr{D}^{\mu}[\mathscr{A}] \mathscr{C} \times i \bar{\mathscr{C}} + \mathscr{A}^{\mu} \times \mathscr{N} \nonumber\\
		& + J^{\mu} .
	\end{align}
The  conserved Noether charge $Q^A := \int d^3x J^{\mu=0,A}_{\rm color}$ obtained from the color current $\mathscr{J}^0_{\rm color}$ is called the \textbf{color charge}
and is equal to the generator of the color rotation.
Note that $\mathscr{J}^{\mu}_{\rm color}$ has the same expression as the massless case, irrespective of $M=0$ or $M\ne0$.

Remember that
	\begin{align}
		i {\boldsymbol \delta}' \bar{{\boldsymbol \delta}}' \mathscr{A}_{\mu}
		&= {\boldsymbol \delta}' (\mathscr{D}_{\mu}[\mathscr{A}] i \bar{\mathscr{C}}) \nonumber\\
		&= {\boldsymbol \delta}' (\partial_{\mu} i \bar{\mathscr{C}} + g \mathscr{A}_{\mu} \times i \bar{\mathscr{C}}) \nonumber\\
		&= - \partial^{\mu} \mathscr{N} + g \mathscr{D}_{\mu}[\mathscr{A}] \mathscr{C} \times i \bar{\mathscr{C}}
		- g \mathscr{A}_{\mu} \times \mathscr{N} ,
	\end{align}
which has the same form for $M=0$ and $M \ne0$. Therefore, we have
	\begin{align}
	&	g \mathscr{J}^{\mu}_{\rm color} + i {\boldsymbol \delta}' \bar{{\boldsymbol \delta}}' \mathscr{A}^{\mu} \nonumber\\
		=& g \mathscr{A}_{\nu} \times \mathscr{F}^{\nu \mu} - \partial^{\mu} \mathscr{N}
		+ gi \partial^{\mu}\bar{\mathscr{C}} \times \mathscr{C} + g J^{\mu} .
	\end{align}
Using this result, we find that the equation of motion for $\mathscr{A}_{\mu}$ is cast into the Maxwell-like form:
	\begin{equation}
		\partial_{\nu} \mathscr{F}^{\nu \mu} + g \mathscr{J}^{\mu}_{\rm color}
		+ i {\boldsymbol \delta}' \bar{{\boldsymbol \delta}}' \mathscr{A}^{\mu} + M^2 \mathscr{A}^{\mu} = 0 .
		\label{YMeq-Maxwell}
	\end{equation}

\section{Defining a massive Yang-Mills field}

We require the following properties to construct a non-Abelian massive spin-one vector boson field $\mathscr{K}_{\mu}(x)$ in a non-perturbative way:
\renewcommand{\theenumi}{\roman{enumi}}
\renewcommand{\labelenumi}{(\theenumi)}
\begin{enumerate}
\item 
$\mathscr{K}_{\mu}$ has the  modified  BRST-invariance (off mass shell): 
	\begin{equation}
		{\boldsymbol \delta}' \mathscr{K}_{\mu} = 0 .
	\end{equation}

\item 
$\mathscr{K}_{\mu}$ is divergenceless (on mass shell): 
	\begin{equation}
		\partial^{\mu} \mathscr{K}_{\mu} = 0 .
	\end{equation}

\item 
$\mathscr{K}_{\mu}$ obeys the adjoint transformation under the color rotation:
	\begin{equation}
		\mathscr{K}_{\mu}(x) \to U \mathscr{K}_{\mu}(x) U^{-1} , \quad U = \exp[i \varepsilon^A Q^A] ,
	\end{equation}	
	which has the infinitesimal version: 
	\begin{equation}
		\delta \mathscr{K}_{\mu}(x) = \varepsilon \times \mathscr{K}_{\mu}(x) . 
	\end{equation}
\end{enumerate}
The field $\mathscr{K}_\mu$ is identified with the non-Abelian version of the physical massive vector field with spin one, as assured by the above properties.
Here (i) guarantees that $\mathscr{K}_{\mu}$ belong  to the physical field creating a physical state with positive norm.
(ii)  guarantees that $\mathscr{K}_{\mu}$ have the correct degrees of freedom as a massive spin-one particle, i.e., three in the four-dimensional spacetime, i.e., two transverse and one longitudinal modes, excluding one scalar mode. 
(iii) guarantees that $\mathscr{K}_{\mu}$ obey the same transformation rule as that of the original gauge field $\mathscr{A}_{\mu}$ .

We find such a field $\mathscr{K}_{\mu}$ is obtained by a nonlinear but local transformation from the original fields  $\mathscr{A}_\mu$,  $\mathscr{C}$,  $\bar{\mathscr{C}}$ and $\mathscr{N}$ of the CF model:
\begin{align}
 \mathscr{K}_\mu :=&  \mathscr{A}_\mu - M^{-2} \partial_\mu \mathscr{N} 
- gM^{-2} \mathscr{A}_\mu \times \mathscr{N} 
\nonumber\\
&+ gM^{-2}  \partial_\mu \mathscr{C} \times i\bar{\mathscr{C}} 
+ g^2 M^{-2} (\mathscr{A}_\mu \times \mathscr{C}) \times i \bar{\mathscr{C}} .
\label{K}
\end{align}
In the Abelian limit or the lowest order of the coupling constant $g$, $\mathscr{K}_{\mu}$ reduces to the Proca field for massive vector:
	\begin{equation}
		\mathscr{K}_{\mu} \to \mathscr{A}_{\mu} - \frac{1}{M^2} \partial_{\mu} \mathscr{N} := U_{\mu} .
	\end{equation}

The new field $\mathscr{K}_{\mu}$ is converted to a simple form:
	\begin{equation}
		\mathscr{K}_{\mu}(x) = \mathscr{A}_{\mu}(x) + \frac{1}{M^2} i {\boldsymbol \delta}' \bar{{\boldsymbol \delta}}' \mathscr{A}_{\mu}(x) .
		\label{K-def}
	\end{equation}
In fact, the definition (\ref{K-def}) for   $\mathscr{K}_{\mu}$  is equal to (\ref{K}):
	\begin{align}
		\mathscr{K}_{\mu} &= \mathscr{A}_{\mu}
		+ \frac{i}{M^2} {\boldsymbol \delta}' (\partial_{\mu} \bar{\mathscr{C}} + g\mathscr{A}_{\mu}\times \bar{\mathscr{C}}) \nonumber\\
		&= \mathscr{A}_{\mu}
		+ \frac{i}{M^2} (i \partial_{\mu} \mathscr{N} + g\mathscr{D}_{\mu}[\mathscr{A}] \mathscr{C} \times \bar{\mathscr{C}}
		+ g \mathscr{A}_{\mu} \times i \mathscr{N} ) \nonumber\\
		&= \mathscr{A}_{\mu}
		- \frac{1}{M^2} (\partial_{\mu} \mathscr{N} + g \mathscr{A}_{\mu} \times \mathscr{N}
		- g\mathscr{D}_{\mu}[\mathscr{A}] \mathscr{C} \times i\bar{\mathscr{C}} ) \nonumber\\
		&= \mathscr{A}_{\mu}
		- \frac{1}{M^2} \partial_{\mu} \mathscr{N} - \frac{g}{M^2} \mathscr{A}_{\mu} \times \mathscr{N}
		+ \frac{g}{M^2} \partial_{\mu}\mathscr{C} \times i\bar{\mathscr{C}}
\nonumber\\
		& \quad + \frac{g^2}{M^2} (\mathscr{A}_{\mu} \times \mathscr{C}) \times i \bar{\mathscr{C}} .
	\label{K_mu1}
	\end{align}

The above properties required for the field $\mathscr{K}_{\mu}$ are checked as follows.
\renewcommand{\theenumi}{\roman{enumi}}
\renewcommand{\labelenumi}{(\theenumi)}
\begin{enumerate}
\item 
This is because
	\begin{align}
	&	{\boldsymbol \delta}' \mathscr{K}_{\mu}  \nonumber\\
&= {\boldsymbol \delta}' \mathscr{A}_{\mu}
		+ \frac{i}{M^2} {\boldsymbol \delta}' {\boldsymbol \delta}' (\partial_{\mu} \bar{\mathscr{C}} + g\mathscr{A}_{\mu}\times \bar{\mathscr{C}}) \nonumber\\
		&= {\boldsymbol \delta}' \mathscr{A}_{\mu} + \frac{i}{M^2} {\boldsymbol \delta}'(\partial_{\mu} {\boldsymbol \delta}' \bar{\mathscr{C}}
		+ g {\boldsymbol \delta}' \mathscr{A}_{\mu}\times \bar{\mathscr{C}} + g \mathscr{A}_{\mu}\times {\boldsymbol \delta}'\bar{\mathscr{C}}) \nonumber\\
		&= {\boldsymbol \delta}' \mathscr{A}_{\mu} + \frac{i}{M^2} (\partial_{\mu} {\boldsymbol \delta}^{'2} \bar{\mathscr{C}}
		+ g {\boldsymbol \delta}^{'2} \mathscr{A}_{\mu}\times \bar{\mathscr{C}} - g {\boldsymbol \delta}' \mathscr{A}_{\mu}\times {\boldsymbol \delta}' \bar{\mathscr{C}} \nonumber\\
		& \hspace{7pc} + g {\boldsymbol \delta}' \mathscr{A}_{\mu}\times {\boldsymbol \delta}' \bar{\mathscr{C}} + g \mathscr{A}_{\mu}\times {\boldsymbol \delta}^{'2} \bar{\mathscr{C}}) \nonumber\\
		&= {\boldsymbol \delta}' \mathscr{A}_{\mu} + \frac{i}{M^2} \mathscr{D}_{\mu}[\mathscr{A}] {\boldsymbol \delta}^{'2} \bar{\mathscr{C}} = 0 ,
	\end{align}
where we have used 
$\bar{{\boldsymbol \delta}}' \mathscr{A}_{\mu}=\mathscr{D}_{\mu}[\mathscr{A}] \bar{\mathscr{C}}$,
${\boldsymbol \delta}^{'2} \mathscr{A}_{\mu} = 0$,   ${\boldsymbol \delta}^{'2} \bar{\mathscr{C}} = iM^2 \mathscr{C}$, and
${\boldsymbol \delta}' \mathscr{A}_{\mu}=\mathscr{D}_{\mu}[\mathscr{A}] \mathscr{C}$.

\item 
The  field equation is obtained:  
	\begin{align}
		\frac{\delta \mathscr{L}^{\rm tot}}{\delta \mathscr{A}^{\mu}}
		=& \mathscr{D}^{\nu}[\mathscr{A}] \mathscr{F}_{\nu \mu} - \partial_{\mu} \mathscr{N} + gi \partial_{\mu} \bar{\mathscr{C}} \times \mathscr{C}
	\nonumber\\&
	+ M^2 \mathscr{A}_{\mu} + g J_{\mu} = 0 , 
		\label{field-eq2}
	\end{align}
	where we have used the left derivative and defined the matter current $J_{\mu}$ by
	\begin{align}
		J_{\mu}^A  =  -i (T^A \varphi)_a \frac{\partial \mathscr{L}^{\rm tot}}{\partial (\partial^{\mu} \varphi_a)} .
	\end{align}
The Noether current associated with  color symmetry, which is conserved  in the sense $\partial_{\mu} \mathscr{J}^{\mu}_{\rm color}=0$, is given by 
	\begin{align}
		\mathscr{J}^{\mu}_{\rm color}
		 =& \mathscr{A}_{\nu} \times \mathscr{F}^{\nu \mu} + i \partial^{\mu}\bar{\mathscr{C}} \times \mathscr{C}
		- \mathscr{D}^{\mu}[\mathscr{A}] \mathscr{C} \times i \bar{\mathscr{C}} 
\nonumber\\& 
+ \mathscr{A}^{\mu} \times \mathscr{N}  + J^{\mu} .
	\end{align}
Using this result, the equation of motion for $\mathscr{A}_{\mu}$ is cast into the Maxwell-like form:
	\begin{equation}
		\partial_{\nu} \mathscr{F}^{\nu \mu} + g \mathscr{J}^{\mu}_{\rm color}
		+ i {\boldsymbol \delta}' \bar{{\boldsymbol \delta}}' \mathscr{A}^{\mu} + M^2 \mathscr{A}^{\mu} = 0 .
		\label{YMeq-Maxwell2}
	\end{equation}
This indeed leads to
	\begin{align}
		\partial^{\mu} \mathscr{K}_{\mu}
		&= \partial^{\mu} \left( \mathscr{A}_{\mu} + \frac{1}{M^2} i {\boldsymbol \delta}' \bar{{\boldsymbol \delta}}' \mathscr{A}_{\mu} \right) \nonumber\\
		&= \frac{-1}{M^2} \partial^{\mu} (\partial^{\nu} \mathscr{F}_{\nu \mu} + g \mathscr{J}_{\mu}^{\rm color}) \nonumber\\
		&= \frac{-1}{M^2} (\partial^{\mu} \partial^{\nu} \mathscr{F}_{\nu \mu} + g \partial^{\mu} \mathscr{J}_{\mu}^{\rm color}) = 0 ,
	\end{align}
where we have used $\partial^{\mu} \partial^{\nu} \mathscr{F}_{\nu \mu}= - \partial^{\mu} \partial^{\nu} \mathscr{F}_{\mu \nu}=0$ and $\partial^{\mu} \mathscr{J}_{\mu}^{\rm color} = 0$.

\item 
	This is trivial from   the Lie-algebra form:
	\begin{align}
		\mathscr{K}_{\mu} =& \mathscr{A}_{\mu}
		- \frac{1}{M^2} \partial_{\mu} \mathscr{N} + i \frac{g}{M^2} [ \mathscr{A}_{\mu} , \mathscr{N} ]
		- i \frac{g}{M^2} [ \partial_{\mu}\mathscr{C} , i\bar{\mathscr{C}} ]
		\nonumber\\&
		- \frac{g^2}{M^2} [ [ \mathscr{A}_{\mu} , \mathscr{C} ] , i \bar{\mathscr{C}} ] .
	\end{align}

\end{enumerate}


As an immediate application of the above result, we can construct a mass term which is invariant simultaneously under the modified BRST transformation,  Lorentz transformation, and  color rotation:
	\begin{equation}
		\frac{1}{2} M^2 \mathscr{K}_{\mu}(x) \cdot \mathscr{K}^{\mu}(x) .
	\end{equation}
This can be useful as a regularization  scheme for avoiding infrared divergences in non-Abelian gauge theories. 
Moreover, we can obtain a dimension-two condensate which is modified  BRST invariant, Lorentz invariant, and color-singlet: 
	\begin{equation}
		\langle \mathscr{K}_{\mu}(x) \cdot \mathscr{K}^{\mu}(x) \rangle .
	\end{equation}
	This dimension-two condensate is off-shell (modified) BRST invariant and  should be compared with the dimension-two condensate proposed in \cite{Kondo01,KMSI02}:
	\begin{equation}
		\langle \frac12 \mathscr{A}_{\mu}(x) \cdot \mathscr{A}^{\mu}(x) + \beta  \mathscr{C}(x) \cdot \mathscr{\bar C}(x) \rangle ,
	\end{equation}
which is only on-shell BRST invariant.

The original CF Lagrangian $\mathscr{L}_{\rm mYM}^{\rm tot}[\mathscr{A}_\mu,\mathscr{C},\bar{\mathscr{C}},\mathscr{N}]$ is written in terms of $\mathscr{A}_\mu, \mathscr{C}, \bar{\mathscr{C}}$ and $\mathscr{N}$.
The new theory is specified by $\mathscr{L}_{\rm mYM}^{\rm tot}[\mathscr{K}_\mu,\mathscr{C},\bar{\mathscr{C}},\mathscr{N}]$ written in terms of $\mathscr{K}_\mu, \mathscr{C}, \bar{\mathscr{C}}$ and $\mathscr{N}$ with the symmetry:
\begin{align}   
\begin{cases}
			{\boldsymbol \delta}^\prime \mathscr{K}_{\mu}(x) =  0
\\
			{\boldsymbol \delta}^\prime \mathscr{C}(x) 
=  -\frac{g}{2} \mathscr{C}(x) \times \mathscr{C}(x)  
\\
			{\boldsymbol \delta}^\prime \bar{\mathscr{C}}(x) =  i \mathscr{N}(x) 
\\
			{\boldsymbol \delta}^\prime \mathscr{N}(x) =  M^2 \mathscr{C}(x) 
\end{cases} 
 .
 \label{Sym}
\end{align}

The proposed model opens a path of resolving the long-standing problem of reconciling physical unitarity with renormalizability without Higgs fields.
The physical unitarity of the CF model will be discussed in  forthcoming papers \cite{Kondo12}.

\section{Remarks}

\subsection{On the choice of the model Lagrangian}

Both the lattice studies and the Schwinger-Dyson equation studies use the conventional QCD action, with none of the terms proposed by this paper or by other gluon-mass modelers, and they all obtain a result suggesting the existence of a  nonperturbative gluon mass. 
It is certainly possible to model the gluon mass result in a possibly nonrenormalizable effective action that is to be used at a classical level (e.g., Ref. \cite{KG67}) that has none of the complications pointed out by us for our own proposal. 
Some readers may ask: 
Are we suggesting that one abandons the standard QCD Lagrangian and use ours? 
If our model is just as an effective action, why must it be renormalizable, usually not a requirement for an effective action? 
Therefore, we would like to make some   comments about these questions. 

\begin{enumerate}
\item[(i)] 
We regard our model just as an effective theory in the sense that it is useful to discuss some aspects on confinement issue relevant in the low-energy regime.  
Therefore, our model as  an effective action is  not to be used at high energies where not only ``unitarity'' is violated, but also ``physical unitarity'', which requires a cutoff at a point (depending on a parameter $\beta$ of the CF model that is physical but not determined by the CF theory).

\item[(ii)]
 In general, the effective model to be valid in low-energy regime needs not  be renormalizable. 
But the renormalizability is a good property in performing calculations in the quantum field theory, and therefore, we have no reason to abandon the renormalizability of the model, if it is maintained. 
In this paper we have adopted the CF model which reduces in the massless limit to the Yang-Mills theory in the Lorenz gauge including the Landau gauge.  
The CF model happens to be renormalizable in this choice of the Lagrangian, which has the same field contents as those in the Yang-Mills theory.  
But remember that even if the renormalizability holds in one gauge, it can be easily lost by taking another gauge. 
This means that this kind of  renormalizability is of some technical character, rather than reflecting the true physics to be described by the model, which should be independent of the gauge choice.  
In our opinion, on the other hand, the physical unitarity must hold in any other gauge choice if it holds for a choice of gauge, in marked contrast to the renormalizability.  

\item[(iii)]
 We expect that our model incorporates some important aspects for understanding the decoupling solution even with a simpler Lagrangian than other approaches, e.g., coming from dynamically generated gluon mass due to QCD dynamics (leading to the momentum-dependent approximate mass function after solving the nonlinear equations numerically \cite{Papa}) and coming from restricting the integration region to the Gribov region or the fundamental modular region  (leading to more complicated refined Gribov-Zwanziger approach \cite{GZ,rGZ}), since we do not yet know how to incorporate these aspects exactly. 

\item[(iv)]
 We hope that our study gives a clue which enables us to extract two transverse modes (corresponding to the physical modes in the massless case) from our model by imposing suitable constraints.  If this is successful, we will be able to give a novel connection between scaling and decoupling.  
\end{enumerate}

\subsection{On the meaning of confinement}

We remember two papers of Kugo and Ojima \cite{KO79,KO78} on color confinement: The first paper \cite{KO79} makes the claim that color confinement requires a certain function $u(p)$  called  the Kugo-Ojima function to have the value $u(0)=-1$, and the second paper \cite{KO78} introduces the quartet mechanism enforcing ``physical unitarity'', provided that the BRST operator is nilpotent.
The value $u(0)=-1$ makes the ghost propagator more singular at zero momentum than it is in perturbation theory. 
However, the decoupling solution in Landau gauge does not fulfill this criterion, and the ghost propagator is not singular at zero momentum. 
In this paper and subsequent papers \cite{Kondo12} the gluon is massive and the ghost propagator is not singular. 
In addition, the BRST operator introduced in these papers  is not nilpotent, as pointed out. 
Therefore, we would like to make some qualitative comments about how there can still be confinement in our model in view of the results of Ref. \cite{KO79}, if the ghost propagator is not singular for generic values of $\beta$, and what significance there is to the physical parameter $\beta$.

\begin{enumerate} 
\item[(i)]
 We consider the Kugo-Ojima criterion is a sufficient condition for color confinement, but it is not a necessary condition for color confinement.  Therefore, the dissatisfaction of the Kugo-Ojima criterion does not necessarily mean the failure of confinement.  
In fact, if we take into account the Gribov copies, the Kugo-Ojima formulation based on the usual BRST method loses its foundation, since it is known that the usual BRST symmetry with nilpotency no longer exists after removing the Gribov copies \cite{GZ,rGZ} and that the nilpotent BRST transformation can be constructed by modifying the usual BRST only when nonlocality of the transformation is allowed \cite{Sorella09,Kondo09b}.  

\item[(ii)]
 It is well known that the existence of the nilpotent BRST symmetry leads to physical unitarity \cite{KO79,KO78}.  But, there is no general proof that the loss of nilpotency immediately yields the violation of physical unitarity.  Therefore, even in the absence of nilpotency, we have still room to find the other way of proving physical unitarity, as we will discuss in subsequent papers \cite{Kondo12}.  

\item[(iii)]
 There could exist other criteria for color confinement.  This viewpoint is supported by the result of Braun, Gies and Pawlowski \cite{BGP10}, which has shown that both decoupling and scaling solution exhibit quark confinement in the sense that the Polyakov loop average is zero, independently of the gauge choice. This result suggests existence of a criterion for color confinement which is applicable to both solutions.  

\item[(iv)]
 For the significance of the parameter $\beta$ as a physical one,  we agree that the choice of the beta determines the content of the theory \cite{Lavrov12}.  In our opinion, the choice of $\beta =0$ will be preferred to explain confinement, since for confinement to occur there must exist a mode which carries the confining force to the long distance.  This seems to be possible only when $\beta =0$, i.e., the ghosts are massless; otherwise both gluons and ghosts become massive in the CF model and the confining force becomes short range.  According to Ref.\cite{TW11},  the simplest case $\beta =0$ well reproduces the decoupling solution obtained in the numerical simulation on a lattice. 

\end{enumerate}

{\it Acknowledgements}\ ---
The author would like to thank the referee for giving valuable comments. 
He would like to thank Ruggero Ferrari for discussions and kind hospitality to him in the stay of Milano in the middle of June 2012, and also Jan Pawlowski for giving  constructive  comments on the issues discussed in this paper and kind hospitality to him in the stay of Heidelberg in the beginning of October 2012. 
This work is  supported by Grant-in-Aid for Scientific Research (C)  24540252 from Japan Society for the Promotion of Science (JSPS).

\appendix
\section{Invariance of the path-integration measure }

In the massive Yang-Mills theory, 
the path-integral integration measure defined by  
	\begin{align}
		&\mathcal{D} \mathscr{A} \mathcal{D} \mathscr{C} \mathcal{D} \bar{\mathscr{C}} \mathcal{D} \mathscr{N} 
		\nonumber\\
   =& \prod_{x, \mu, A} d\mathscr{A}_{\mu}^A(x) \prod_{x, A} d\mathscr{C}^A(x)
		\prod_{x, A} d\bar{\mathscr{C}}^A(x) \prod_{x, A} d\mathscr{N}^A(x) 
	\end{align}
is invariant under the modified BRST transformation.
This fact is shown as follows.
For the change of integration variables $\Phi(x) \to \Phi'(x) =\Phi(x)+ \lambda {\boldsymbol \delta}' \Phi(x)$, the integration measure is transformed as
	\begin{align}
	&	 \mathcal{D} \mathscr{A}' \mathcal{D} \mathscr{C}' \mathcal{D} \bar{\mathscr{C}}' \mathcal{D} \mathscr{N}'  
		=   J \mathcal{D} \mathscr{A} \mathcal{D} \mathscr{C} \mathcal{D} \bar{\mathscr{C}} \mathcal{D} \mathscr{N} ,
\nonumber\\
  J =&  {\rm Det}
		\left[ \begin{array}{@{\,}cccc@{\,}}
			\frac{\delta \mathscr{A}_{\mu}^{'A}(x)}{\delta \mathscr{A}_{\nu}^B(y)} & \frac{\delta \mathscr{A}_{\mu}^{'A}(x)}{\delta \mathscr{C}^B(y)} &
			\frac{\delta \mathscr{A}_{\mu}^{'A}(x)}{\delta \bar{\mathscr{C}}^B(y)} & \frac{\delta \mathscr{A}_{\mu}^{'A}(x)}{\delta \mathscr{N}^B(y)} \\ 
			\frac{\delta \mathscr{C}^{'A}(x)}{\delta \mathscr{A}_{\nu}^B(y)} & \frac{\delta \mathscr{C}^{'A}(x)}{\delta \mathscr{C}^B(y)} &
			\frac{\delta \mathscr{C}^{'A}(x)}{\delta \bar{\mathscr{C}}^B(y)} & \frac{\delta \mathscr{C}^{'A}(x)}{\delta \mathscr{N}^B(y)} \\ 
			\frac{\delta \bar{\mathscr{C}}^{'A}(x)}{\delta \mathscr{A}_{\nu}^B(y)} & \frac{\delta \bar{\mathscr{C}}^{'A}(x)}{\delta \mathscr{C}^B(y)} &
			\frac{\delta \bar{\mathscr{C}}^{'A}(x)}{\delta \bar{\mathscr{C}}^B(y)} & \frac{\delta \bar{\mathscr{C}}^{'A}(x)}{\delta \mathscr{N}^B(y)} \\ 
			\frac{\delta \mathscr{N}^{'A}(x)}{\delta \mathscr{A}_{\nu}^B(y)} & \frac{\delta \mathscr{N}^{'A}(x)}{\delta \mathscr{C}^B(y)} &
			\frac{\delta \mathscr{N}^{'A}(x)}{\delta \bar{\mathscr{C}}^B(y)} & \frac{\delta \mathscr{N}^{'A}(x)}{\delta \mathscr{N}^B(y)} \\ 
		\end{array} \right] \nonumber\\
	\end{align}
The Jacobian $J$ is calculated as
\footnotesize
	\begin{align}
	J	=& 
 {\rm Det}
		\left[ \begin{array}{@{\,}cccc@{\,}}
			\delta_{\mu \nu} (\delta^{AB} + \lambda g f^{ABC} \mathscr{C}^C) & \lambda \mathscr{D}^{AB}_{\mu}[\mathscr{A}] & 0 & 0 \\
			0 & \delta^{AB} - \lambda g f^{ABC} \mathscr{C}^C & 0 & 0 \\
			0 & 0 & \delta^{AB} & \lambda i \delta^{AB} \\
			0 & \lambda M^2 \delta_{AB} & 0 & \delta_{AB}
		\end{array} \right] \nonumber\\	& \times
\delta^D(x-y) \nonumber\\		 
		=& 
 {\rm Det}
		\left[ \begin{array}{@{\,}ccc@{\,}}
			\delta_{\mu \nu} (\delta^{AB} + \lambda g f^{ABC} \mathscr{C}^C) & \lambda \mathscr{D}^{AB}_{\mu}[\mathscr{A}] & 0 \\
			0 & \delta^{AB} - \lambda g f^{ABC} \mathscr{C}^C & 0 \\
			0 & 0 & \delta^{AB} 
		\end{array} \right] \nonumber\\	& \times
\delta^D(x-y) \nonumber\\	
		& \quad + (\lambda M^2) \cdot {\rm Det}
		\left[ \begin{array}{@{\,}ccc@{\,}}
			\delta_{\mu \nu} (\delta^{AB} + \lambda g f^{ABC} \mathscr{C}^C) & 0 & 0 \\
			0 & 0 & 0 \\
			0 & \delta^{AB} & \lambda i \delta^{AB}
		\end{array} \right] \nonumber\\	& \times
\delta^D(x-y) \nonumber\\	
		=& 
		{\rm Det}[\delta_{\mu \nu} \delta^D(x-y) (\delta^{AB} + \lambda g f^{ABC} \mathscr{C}^C)]
\nonumber\\		 
		 & \times {\rm Det}[\delta^D(x-y)(\delta^{AB} - \lambda g f^{ABC} \mathscr{C}^C)] \nonumber\\
		=& 
1 .
\label{J-calc}
	\end{align}

\normalsize
\section{$\beta$ dependence}

We have shown that 
$\mathscr{L}_{\rm GF+FP}$ is written as
	\begin{align}
		\mathscr{L}_{\rm GF+FP}  
		 =& i {\boldsymbol \delta}' \bar{{\boldsymbol \delta}}' \left(\frac{1}{2} \mathscr{A}^{\mu} \cdot \mathscr{A}_{\mu}
		+ \frac{\beta}{2} i \bar{\mathscr{C}} \cdot \mathscr{C} \right) - \frac{\beta}{2} M^2 i \bar{\mathscr{C}} \cdot \mathscr{C} ,
	\end{align}
while 
	\begin{equation}
		\mathscr{L}_{m} = \frac{1}{2} M^2 \mathscr{A}_{\mu} \cdot \mathscr{A}^{\mu} + \beta M^2 i \bar{\mathscr{C}} \cdot \mathscr{C} .
	\end{equation}
Thus, we have	
	\begin{align}
	\mathscr{L}_{\rm GF+FP} + \mathscr{L}_{m} 
=& i {\boldsymbol \delta}' \bar{{\boldsymbol \delta}}' \left(\frac{1}{2} \mathscr{A}^{\mu} \cdot \mathscr{A}_{\mu}
		+ \frac{\beta}{2} i \bar{\mathscr{C}} \cdot \mathscr{C} \right) 
		 \nonumber\\& 
+ \frac{\beta}{2} M^2 i \bar{\mathscr{C}} \cdot \mathscr{C}
+ \frac{1}{2} M^2 \mathscr{A}_{\mu} \cdot \mathscr{A}^{\mu}  .
	\end{align}
	
Let $W$ be the generating functional of the connected Green functions defined from the vacuum functional $Z[J]$ with the source $J$  for an operator $\mathscr{O}$ as a functional of $\Phi$: 
	\begin{align}
	  e^{iW[J]} 
:=   Z[J] 
&:=   \int \mathcal{D} \mathscr{A} \mathcal{D} \mathscr{C} \mathcal{D} \bar{\mathscr{C}} \mathcal{D} \mathscr{N} 
\nonumber\\& 
\times \exp \left\{ iS^{\rm tot}_{m\rm{YM}} + i\int d^Dx J(x) \cdot \mathscr{O}(x) \right\} .
	\end{align}
Then the average of the operator $\mathscr{O}$ is obtained from
	\begin{align}
	 \langle \mathscr{O}(x) \rangle 
=   \frac{\delta}{\delta J(x)} W[J] \Big|_{J=0} 
= \frac{1}{i}  \frac{\delta}{\delta J(x)} \ln Z[J] \Big|_{J=0} 
  .
	\end{align}

First, we consider the derivative of $W$ with  respect to $\beta$ given by
	\begin{equation}
	\frac{\partial W[J]}{\partial \beta}
	= \frac{1}{i} \frac{\partial \ln Z[J]}{\partial \beta}
	= \frac{1}{i} Z[J]^{-1} \frac{\partial  Z[J]}{\partial \beta}
	= \Big\langle \frac{\partial S^{\rm tot}_{m\rm{YM}}}{\partial \beta} \Big\rangle_J ,
	\end{equation}
where
	\begin{equation}
	 \frac{\partial S^{\rm tot}_{m\rm{YM}}}{\partial \beta} 
= \int d^Dx \left[ i {\boldsymbol \delta}' \bar{{\boldsymbol \delta}}' \left( \frac{1}{2} i \bar{\mathscr{C}} \cdot \mathscr{C} \right) + \frac{1}{2} M^2 i \bar{\mathscr{C}} \cdot \mathscr{C} \right] .
	\end{equation}

If we adopt the modified Kugo-Ojima subsidiary condition  or the modified BRST invariance of the vacuum:
	\begin{equation}
	 Q_{\rm BRST}' | 0 \rangle = 0 ,
	\end{equation}	
	we find the $\beta$ dependence of $W[J]$:
	\begin{equation}
	\frac{\partial W[J]}{\partial \beta}
	= \int d^Dx \frac{1}{2} M^2 \langle  i \bar{\mathscr{C}}(x) \cdot \mathscr{C}(x)  \rangle_J \not= 0
	\end{equation}
Therefore, for $M \not=0$,  $W[J]$ depends on the parameter $\beta$.	
This result should be compared with the $M=0$ case, in which $\beta$ is a gauge-fixing parameter and hence  $W[J]$ should not depend on $\beta$. In the $M=0$ case, any choice of $\beta$ gives the same  $W[J]$. 
However, this is not the case for $M \not=0$.

Next, we consider the average of the operator $\mathscr{O}$:
	\begin{align}
	 \langle \mathscr{O}(x) \rangle 
=    Z^{-1}   \int \mathcal{D} \mathscr{A} \mathcal{D} \mathscr{C} \mathcal{D} \bar{\mathscr{C}} \mathcal{D} \mathscr{N} 
 e^{ iS^{\rm tot}_{m\rm{YM}} } \mathscr{O}(x)
  ,
	\end{align}
where $Z:=Z[J=0]$. 
The $\beta$ derivative  is given by
		\begin{align}
	\frac{\partial}{\partial \beta} \langle \mathscr{O}(x) \rangle 
=  \Big\langle \mathscr{O}(x)  ;  i \frac{\partial S^{\rm tot}_{m\rm{YM}}}{\partial \beta}  \Big\rangle
   ,
	\end{align}
where we have defined the connected expectation value 
$\langle A; B \rangle := \langle A   B \rangle -\langle A \rangle \langle B \rangle $.
Even if the operator $\mathscr{O}$ is modified BRST invariant, i.e., ${\boldsymbol \delta}' \mathscr{O}=0$, the average $ \langle \mathscr{O}(x) \rangle$ depends on $\beta$ for $M \not=0$: 
		\begin{align}
	\frac{\partial}{\partial \beta} \langle \mathscr{O}(x) \rangle 
= & i \frac{1}{2} M^2  \int d^Dy   \Big\langle \mathscr{O}(x)   ;   i \bar{\mathscr{C}}(y) \cdot \mathscr{C}(y) \Big\rangle 
 \not= 0
   .
	\end{align}
The $\beta$ dependence of the CF model was pointed out also in \cite{Lavrov12} by using different arguments.


\end{document}